\documentclass[prb,twocolumn,preprintnumbers,superscriptaddress,longbibliography]{revtex4-2}

\usepackage[dvipdfmx]{graphicx}% Include figure files
\usepackage{dcolumn}% Align table columns on decimal point
\usepackage{bm}% bold math
\usepackage{color}
\usepackage{amssymb}
\usepackage{amsmath}

\begin{document}

\preprint{APS/123-QED}

\title{Electronic inhomogeneity in Cs- and Sb-terminated surfaces of CsV$_3$Sb$_5$ probed by scanning photoemission spectromicroscopy}

\author{T.~Mizokawa}  
\affiliation{Department of Applied Physics, Waseda University, Tokyo 169-8555, Japan}
\author{G.~Tomassucci}   %giovanni.tomassucci@uniroma1.it
\affiliation{Department of Physics, University of Roma "La Sapienza", Roma 00185, Italy}
\author{M.~Hattori}     %mackyhatto@toki.waseda.jp
\affiliation{Department of Applied Physics, Waseda University, Tokyo 169-8555, Japan}
\author{F.~Minati}
\affiliation{Department of Physics, University of Roma "La Sapienza", Roma 00185, Italy}
\author{L.~Tortora}
\affiliation{Department of Physics, University of Roma "La Sapienza", Roma 00185, Italy}
\author{A.~Barinov}
\affiliation{Sincrotrone Trieste S.C.p.A., Area Science Park, 34012 Basovizza, Trieste, Italy}
\author{Z.~Wang} %zhiweiwang@bit.edu.cn
\affiliation{Centre for Quantum Physics, Key Laboratory of Advanced Optoelectronic Quantum Architecture and Measurement (MOE), School of Physics, Beijing Institute of Technology, Beijing 100081, China}
\affiliation{Beijing Key Lab of Nanophotonics and Ultrafine Optoelectronic Systems, Beijing Institute of Technology, Beijing 100081, China}
\author{J.-X.~Yin}   %yinjx@sustech.edu.cn
\affiliation{Department of Physics, Southern University of Science and Technology, Shenzhen, Guangdong 518005, China}
\author{N.~L.~Saini}  
\affiliation{Department of Physics, University of Roma "La Sapienza", Roma 00185, Italy}

\begin{abstract}
Electronic structures of Cs- and Sb-terminated surfaces of a kagome superconductor CsV$_3$Sb$_5$ have been elucidated by means of scanning photoemission microscopy (SPEM). The observed band structure of the Cs-terminated surface is rather close to that of the bulk while that of the Sb-terminated one is substantially modified around K/H point of the Brillouin zone. While the contrast between the Cs- and Sb-terminated regions is reduced below the charge density wave transition temperature, the Sb 5$p$ band of Cs-terminated region exhibits electronic inhomogeneity which slightly increases below it. The inhomogeneity of the Sb 5$p$ band would be related to disorders of the out-of-plane Sb and relevant for the band folding along $\Gamma$-A with the charge density wave.
The SPEM results suggest that the less inhomogeneous Cs termination is more suitable for interface of kagome superconductors. However, the inhomogeneity of Cs termination, which is significant at $\Gamma$/A, noticeable at K/H, and negligible at M/L, is expected to affect the Sb 5$p$-V 3$d$ hybridization at the interface.
\end{abstract}

\maketitle

\section{Introduction}
Since the discovery of charge density wave \cite{Ortiz2019} and superconductivity \cite{Ortiz2020} in $A$V$_3$Sb$_5$ ($A$ = K, Rb, Cs) by Ortiz et al., the interplay between these two macroscopic phases on the V kagome lattice have been attracting great interest in the condensed matter physics community \cite{Yin2022,Wilson2024,Zhong2024}. CsV$_3$Sb$_5$ and KV$_3$Sb$_5$ exhibit $2\times 2\times 2$ lattice distortion below 94 K and 80 K \cite{Liang2021,Li2021,Liu2021} due to charge and/or bond modulation of the V 3$d$ electrons on the kagome lattice. Incidentally, an exotic density wave order with time reversal symmetry breaking is identified in KV$_3$Sb$_5$ \cite{Jiang2021,Ortiz2021,Mielke2022}.
Superconducting transition temperature $T_c$ of CsV$_3$Sb$_5$ and KV$_3$Sb$_5$ are about 2.5 K and 0.9 K, respectively \cite{Ortiz2020,Ortiz2021}. $T_c$ of
CsV$_3$Sb$_5$ exhibits double superconducting domes as a function pressure and the maximum $T_c$ is about 8 K around 2 GPa \cite{Chen2021}. The complicated behaviors of the charge density wave and the superconductivity suggest competition between multiple electronic and lattice instabilities.

\begin{figure}[t]
\centering
\includegraphics[width=0.48\textwidth,pagebox=cropbox,clip]{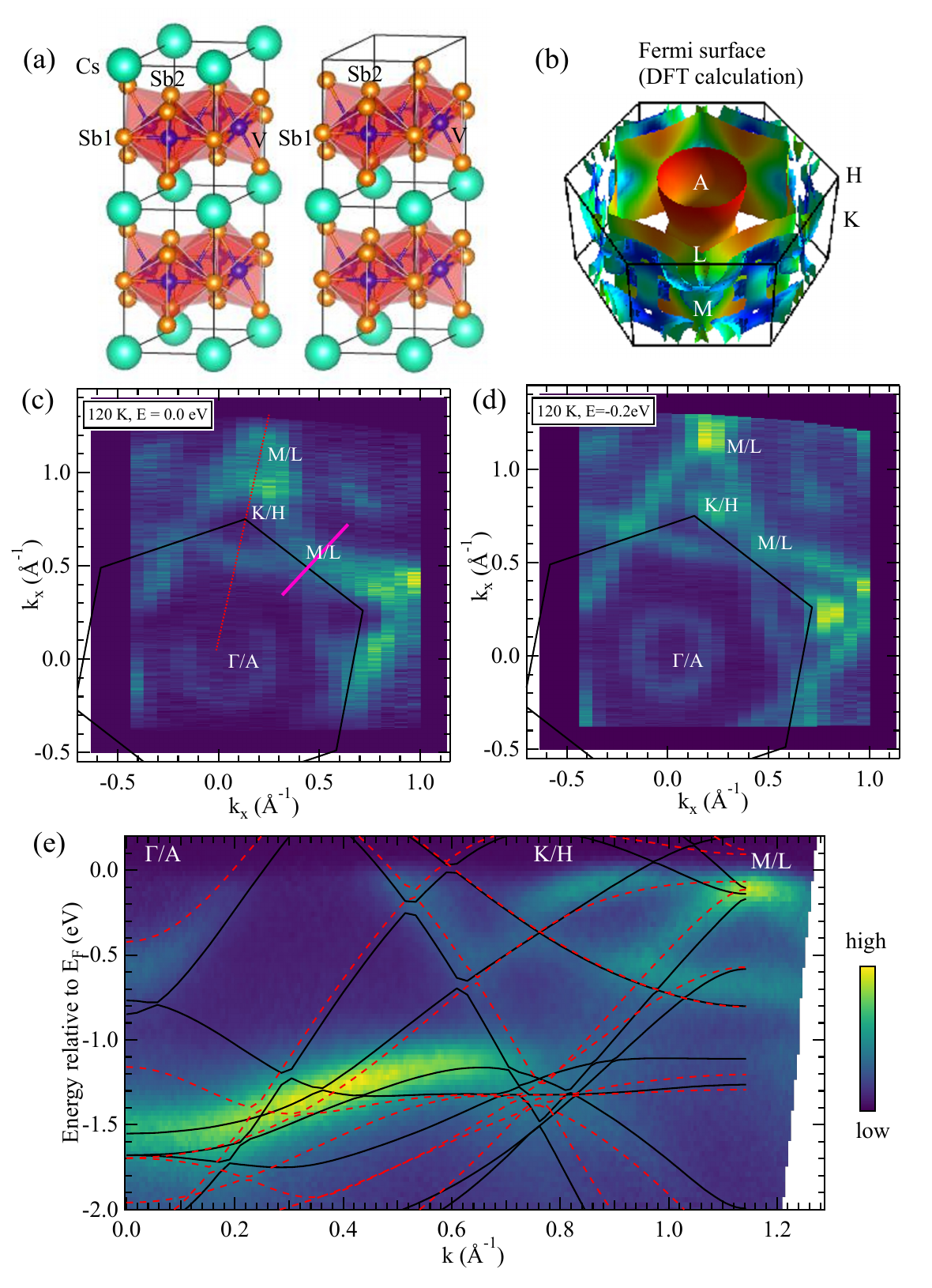}
\caption{(Color online)
(a) Crystal structure of CsV$_3$Sb$_5$ for Cs-terminated (left) and Sb-terminated (right) surfaces. (b) Bulk Brillouin zone and calculated Fermi surfaces.
(c) Fermi surface map taken at 120 K for CsV$_3$Sb$_5$ (position A indicated in Fig. 2). The hexagonal lines indicate the Brillouin zone boundaries. (d) Constant energy map at -0.2 eV taken at 120 K. (e) Band dispersions along the $\Gamma$-K-M or A-H-L cut indicated by the dotted line in (a). Calculated band dispersions along $\Gamma$-K-M and A-H-L are shown by the dashed and solid curves, respectively.
}
\end{figure}

The layered crystal structure of CsV$_3$Sb$_5$ is illustrated in Fig. 1(a). Cs and V$_3$Sb$_5$ layers are stacked along the c-axis. The cleaved surface can be terminated at the Cs layer or the V$_3$Sb$_5$ layer. In the V$_3$Sb$_5$ layer, each V atom is coordinated by two in-plane Sb (Sb1) and four out-of-plane Sb (Sb2).
The Fermi surfaces of CsV$_3$Sb$_5$ are formed by V 3$d$ and Sb 5$p$ bands around K point and $\Gamma$ point of the hexagonal Brillouin zone respectively as revealed by angle-resolved photoemission spectroscopy (ARPES) and are basically consistent with the calculations based on the density functional theory (DFT) \cite{Nakayama2021,Kang2022,Cho2021,Lou2022,Kato2022,Kato2022_2,Hu2022,Zhong2024}. The bulk Brillouin zone and Fermi surfaces calculated by DFT are illustrated in Fig. 1(b). Among the ARPES studies, Nakayama et al. showed that the charge density wave gap is enhanced around M point with van Hove singularities \cite{Nakayama2021}. Kang et al. further revealed that the $k_z$ dependence of the van Hove singularities provides the $2\times 2\times 2$ charge density wave \cite{Kang2022}. The orbital- and momentum-dependent energy gap at the Fermi level in the charge density wave phase does not support the canonical Peierls mechanism due to the electron-lattice interaction. Instead, the van Hove singularities located at M point around 0.1-0.2 eV below the Fermi level are found to show significant changes across the charge density wave transition. The anomalous behaviors of charge density wave have been disclosed by transport and optical measurements \cite{Yu2021,Zhao2021,Chen2021,Zhou2021,ZWang2021,Guo2022}. The superconducting state has been suggested to have complex order parameter with anisotropic s-wave gap \cite{Zheng2022,Zhong2023,Le2024}. The charge density wave is suppressed and the superconductivity is enhanced by pressure \cite{QWang2021,Zhang2021,FYu2021,Qian2021,Zhu2022,Xu2021}. Therefore, the charge density wave competes with the superconductivity. In addition, the nematicity associated with the charge density wave and/or pair density wave has been reported \cite{Xiang2021,Wulferding2022,Song2022}. 
 Theoretically, the origin of the charge density wave and the superconductivity can be assigned to the combination of the electron-electron interaction and the electron-lattice interaction which can be enhanced by the van Hove singularities \cite{Denner2021,Tan2021,Tazai2022,LaBollita2021,Subedi2022}. 

The nematicity and phase competition suggest that electronic inhomogeneity may play vital roles in CsV$_3$Sb$_5$ which can be probed by scanning photoemission spectromicroscopy (SPEM). The SPEM technique has successfully been applied to study phase competition between Mott insulating and superconducting phases in K$_x$Fe$_{2-y}$Se$_2$ \cite{Bendele2014}, nematic electronic structure of FeTe \cite{Mizokawa2016}, and domain-dependent electronic structure in charge density wave phase of IrTe$_2$ \cite{Mizokawa2022}.

In the present work, in order to investigate possible electronic inhomogeneity in the Sb-terminated and Cs-terminated surfaces of the CsV$_3$Sb$_5$ kagome superconductor and its relationship with the charge density wave transition, we have performed SPEM measurements on CsV$_3$Sb$_5$ above and below the transition temperature.
The SPEM results suggest that the less inhomogeneous Cs termination is more suitable for interface of kagome superconductors than the Sb termination. Across the charge density wave transition temperature, the Sb 5$p$ band at Cs termination exhibits increase of electronic inhomogeneity which would be related to disorders of the out-of-plane Sb. The electronic inhomogeneity of the Sb 5$p$ band for Cs termination, significant at $\Gamma$/A, noticeable at K/H, and negligible at M/L, is expected to affect the Sb 5$p$-V 3$d$ hybridization at interface.

\section{Methods}
CsV$_3$Sb$_5$ crystals were grown as reported in the literature \cite{Zhong2023} and characterized by bulk-sensitive hard x-ray photoemission spectroscopy \cite{Takegami2024} excluding possibility of impurities in the bulk. The SPEM measurements were performed at the spectromicroscopy beamline, Elettra synchrotron facility, Trieste, Italy \cite{SPEM}. The single crystals were cleaved at 120 K and measured at 120 K and 60 K. Photons at 27 eV were focused through a Schwarzschild objective in order to obtain a submicron size beam spot. For the present measurements, the total energy resolution was set to about 50 meV and the angle resolution was 1 degree.
The crystal structure is analyzed and illustrated by VESTA \cite{VESTA}. The DFT calculations with generalized gradient approximation (GGA) were performed by Quantum Espresso 6.7 \cite{QE} with pseudo potentials of Cs.pbe-spn-kjpaw\_psl.1.0.0.UPF, V.pbe-spnl-kjpaw\_psl.1.0.0.UPF, and Sb.pbe-n-kjpaw\_psl.1.0.0.UPF. The atomic positions in the cubic unit cell were taken from the literature \cite{Ortiz2019}. The kinetic-energy cutoff and the charge density cutoff were set to 80 Ry and 500 Ry, respectively. We employed the $k$ mesh of 10$\times$10$\times$10 in the Brillouin zone illustrated in Fig. 1(b). The intersections of the bulk Fermi surfaces were constructed by means of FermiSurfer  2.4.0 \cite{FermiSurfer}.

\section{Results and Discussion}

Figures 1(c) and 1(d) show a Fermi surface map and a constant energy map at -0.2 eV for the cleaved surface of CsV$_3$Sb$_5$ at 120 K [for position A1 of Fig. 2(a) corresponding to Cs termination]. The Sb 5$p_z$ Fermi pocket around $\Gamma$/A point and the V 3$d$ states around M/L points are clearly observed consistent with the prediction of DFT \cite{Ortiz2019} and the results of the previous ARPES works \cite{Nakayama2021,Kang2022,Cho2021,Lou2022,Kato2022,Kato2022_2,Hu2022}. Figure 1(e) shows band dispersions along the $\Gamma$-K-M line in Fig. 1(c). The Sb 5$p_z$ band crosses the Fermi level at $k$ = 0.2 \AA$^{-1}$ (around $\Gamma$/A point). The V 3$d$ bands cross the Fermi level around $k$ = 0.6 \AA$^{-1}$ and around $k$ = 0.9 \AA$^{-1}$ (around K/H point). The observed band dispersions are compared with the calculations along $\Gamma$-K-M and A-H-L which are shown by the dashed and solid curves in Fig. 1(e). With the photon energy of 27 eV, the momentum perpendicular to the surface is about 3.0 \AA$^{-1}$ (4.43 $\times$ 2$\pi/c$). Accordingly. the observed band dispersion roughly follows the calculation along A-H-L.

Figures 2(a) and 2(b) show SPEM images and extracted photoemission spectra at $\Gamma$/A for 120 K and 60 K, respectively. The bright (dark) region has higher (lower) photoemission intensity in the SPEM images. At $\Gamma$/A point, the spectral weight near the Fermi level is derived from Sb 5$p$ while the peak at -1.5 eV corresponds to V 3$d$. It is known that the Sb 5$p_z$ band of the in-plane Sb1 site [see Fig. 1(a)]
is accompanied by excess spectral weight on the lower energy side (higher binding energy side) around -0.5 eV in the Cs terminated surface \cite{Kato2022}.
In the top panels, the SPEM images were created for the energy range of the excess spectral weight which is indicated by the shaded area with the upward arrows in the middle panels of Fig. 2. In the SPEM image of Fig. 2(a) at 120 K, the bright (dark) region with the excess spectral weight corresponds to the Cs (Sb) termination. In the Cs terminated region, the photoemission spectra at A1 (red) and A2 (green) exhibit the excess spectral weight around -0.5 eV, which is suppressed in the spectrum at B (blue) in the Sb terminated region. The contrast between the bright and dark regions is reduced at 60 K. In the photoemission spectra at A1, A2 and B, the -0.5 eV spectral weight tends to be suppressed. The suppression of spectral weight around -0.5 eV is stronger at A2 than at A1. This inhomogeneity within the selected Cs-terminated region is clearly seen in Fig. 2(c) which shows the fine scan for the narrow region indicated by the box in Figs. 2(a) and 2(b). This observation suggests that the Sb 5$p$ band at the surface is reconstructed in an inhomogeneous manner across the charge density wave transition. The standard deviation divided by the average value of the Sb 5$p$ photoemission intensity of the selected Cs-terminated region is 0.105 at 120 K and 0.107 (0.110 in the fine scan) at 60 K, suggesting that the inhomogeneity embedded in the cleaved surface increases from 120 K to 60 K only slightly.
In the bottom panels, the SPEM images were created for the energy range of the V 3$d$ band around -1.5 eV, which
is indicated by the shaded area with the downward arrow in the photoemission spectra in the middle panels. The inhomogeneous spectral distribution does not change appreciably across the charge density wave transition.
The standard deviation divided by the average value of the V 3$d$ photoemission intensity of the selected Cs-terminated region is 0.087 at 120 K and 0.080 (0.081 in the fine scan) at 60 K, suggesting that the inhomogeneity embedded in the cleaved surface decreases from 120 K to 60 K for the V 3$d$ band.

The SPEM images in Figs. 3(a) and 3(b) for K/H point were created with the same energy window. The intensity of the V 3$d$ band is largely reduced for the Sb termination compared to that for the Cs termination. At K/H point, the V 3$d$ $xy$/$x^2-y^2$ band hybridized with Sb 5$p_x$/5$p_y$ at out-of-plane Sb2 site are located near the Fermi level and around -1.0 eV \cite{Hu2022}. The intensity of the V 3$d$ $xy$/$x^2-y^2$ bands is relatively reduced in the Sb-terminated surface probably due to the reduction of the hybridization.
The spectral change between 120 K and 60 K at K/H is much smaller than that at $\Gamma$/A. As shown in Fig. 3(c), the inhomogeneity within the Cs-terminated region is smaller than that at $\Gamma$/A. It should be noted that the K/H point for Fig. 3(c) is different from that for Figs. 3(a) and 3(b).

The difference between the Cs- and Sb-terminated surfaces is inappreciable in the SPEM image and the extracted spectra for M/L point as shown in Fig. 4. At M/L point, the V 3$d$ $yz$/$zx$ band near the Fermi level hardly hybridizes with Sb 5$p$ at out-of-plane Sb2 site. Therefore, the difference between the Cs- and Sb-terminations is expected to be small.

\begin{figure}[t]
\centering
\includegraphics[width=0.5\textwidth,pagebox=cropbox,clip]{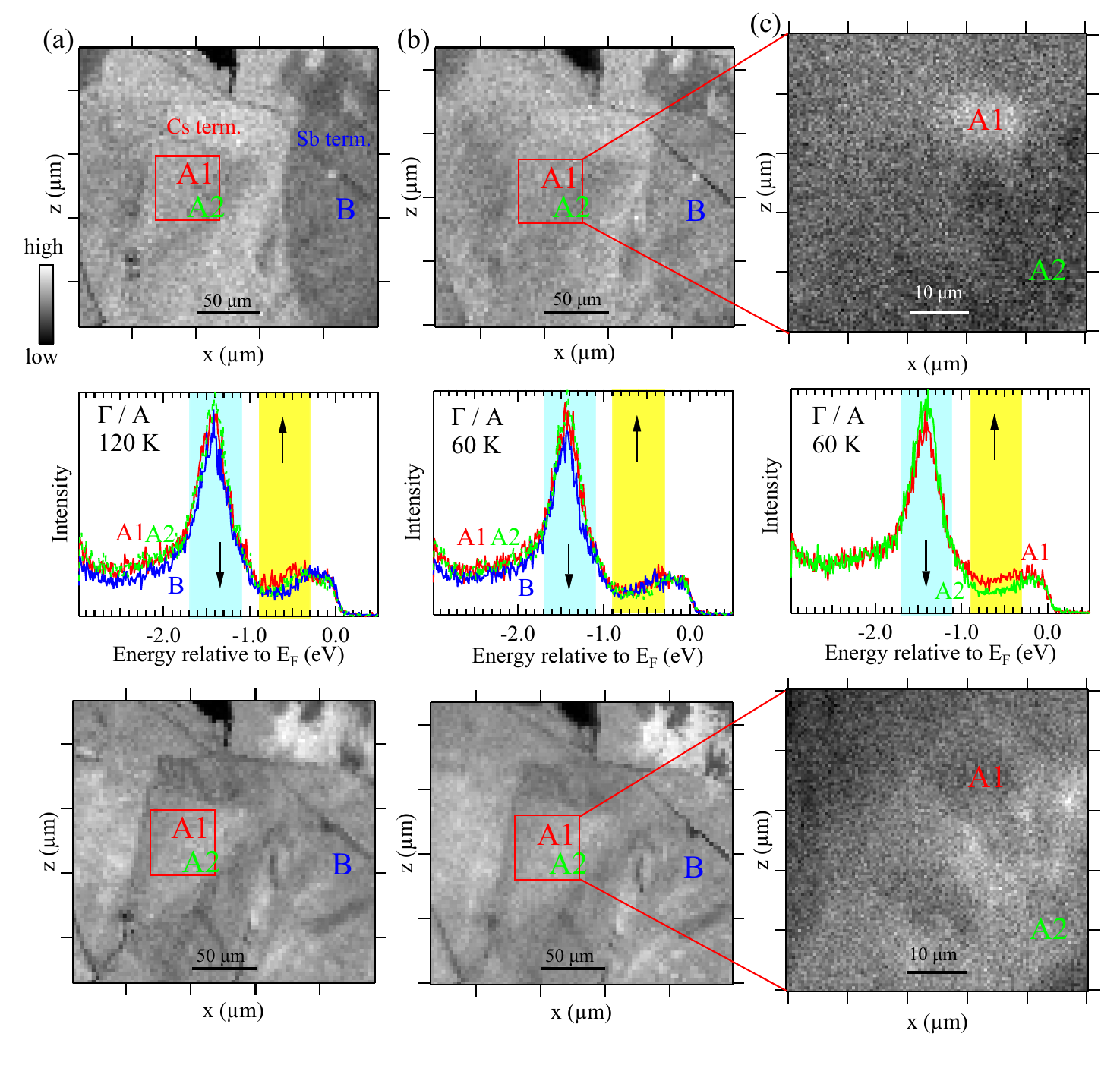}
\caption{(Color online)
SPEM images (top and bottom panels) and photoemission spectra of points A1, A2, and B (middle panels) taken at $\Gamma$/A in the Brillouin zone for (a) 120 K and wide region, (b) 60 K and wide region, and (c) 60 K and narrow region indicated by the box in (a) and (b).
The SPEM images in the top (bottom) panels show spatial distributions of photoemission intensity integrated within the shaded energy window with upward (downward) arrows in the middle panels.
}
\end{figure}

\begin{figure}[t]
\centering
\includegraphics[width=0.5\textwidth,pagebox=cropbox,clip]{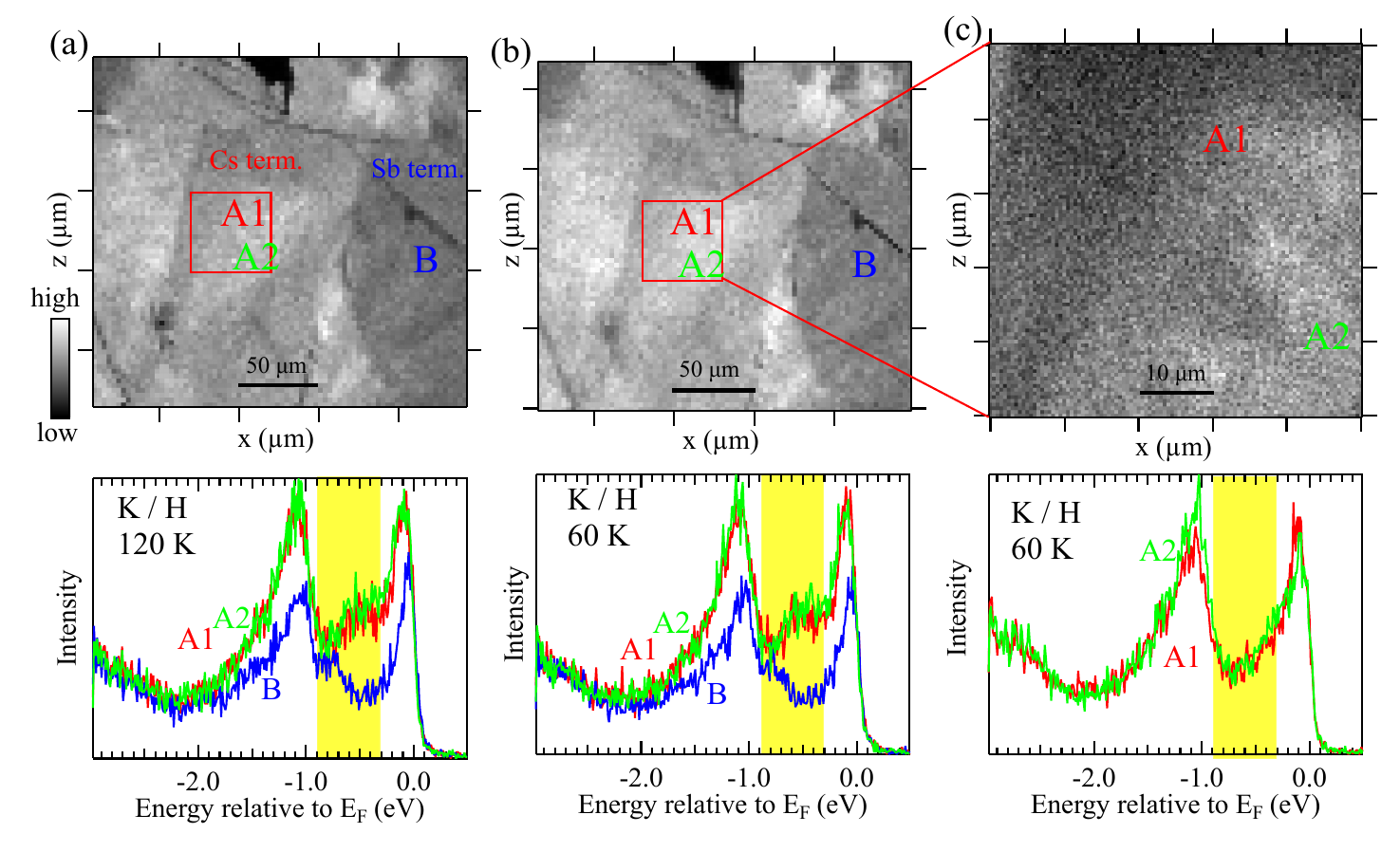}
\caption{(Color online)
SPEM images (upper panels) and photoemission spectra of points A1, A2, and B (lower panels) taken at K/H in the Brillouin zone for (a) 120 K and wide region, (b) 60 K and wide region, and (c) 60 K and narrow region indicated by the box in (a) and (b).
The SPEM images show spatial distributions of photoemission intensity integrated within the energy window shaded in the photoemission spectra.
}
\end{figure}

\begin{figure}[t]
\centering
\includegraphics[width=0.5\textwidth,pagebox=cropbox,clip]{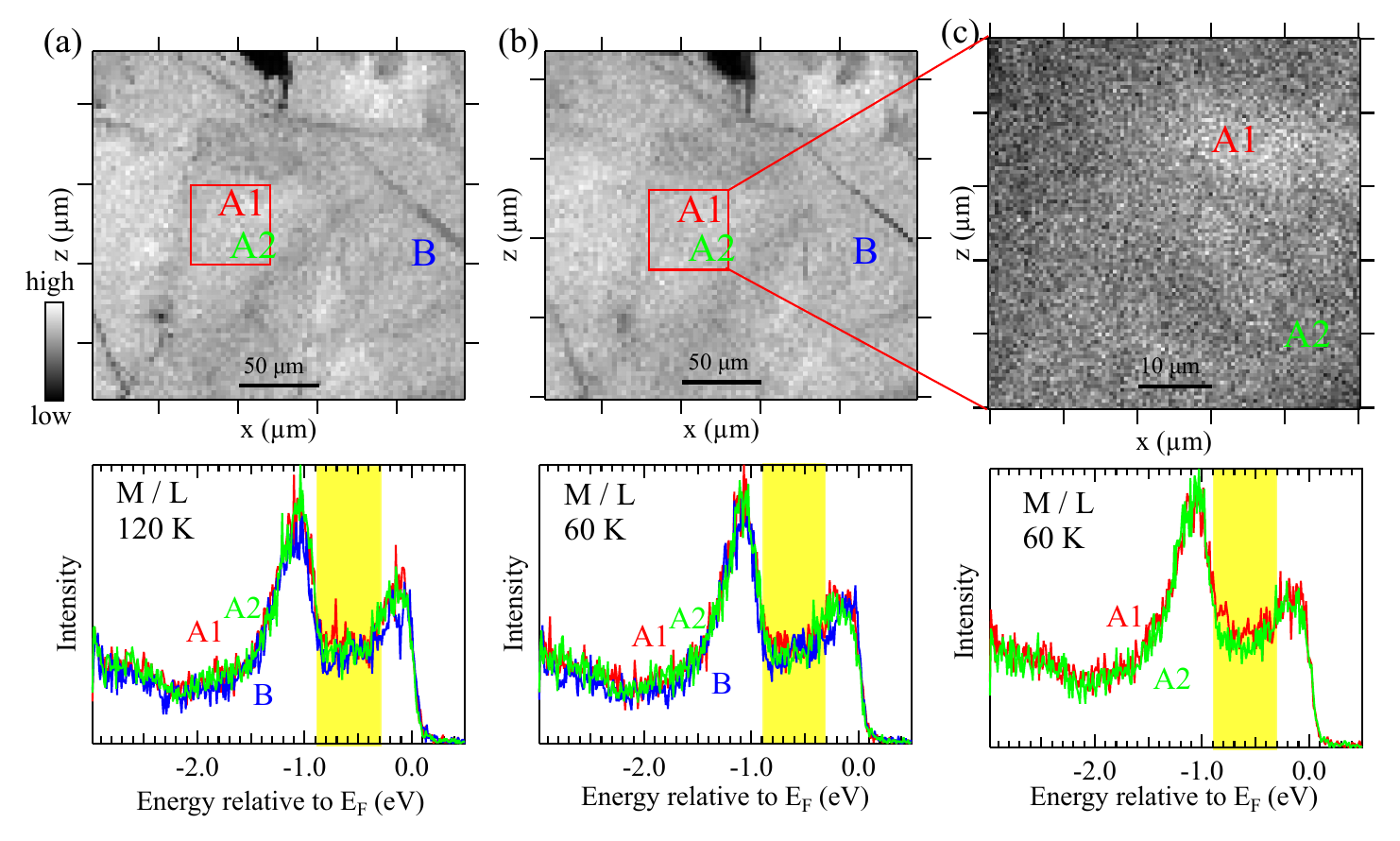}
\caption{(Color online)
SPEM images (upper panels) and photoemission spectra of points A1, A2, and B (lower panels) taken at M/L in the Brillouin zone for (a) 120 K and wide region, (b) 60 K and wide region, and (c) 60 K and narrow region indicated by the box in (a) and (b).
The SPEM images show spatial distributions of photoemission intensity integrated within the energy window shaded in the photoemission spectra.
}
\end{figure}

Figures 5(a) and 5(b) respectively show a Fermi surface map and a constant energy map at -0.2 eV taken at 60 K, which is below the charge density wave transition temperature, around position A1 in the Cs-terminated region.
Figures 5(c) and 5(d) show a Fermi surface map and a constant energy map at -0.2 eV at 60 K around around position B in the Sb-terminated region, respectively. 
In comparison to the map at 120 K shown in Fig. 1(c), the Fermi surface geometry does not change appreciably across the charge density wave transition. At the Fermi level, the spectral weight around K/H point is higher in the Cs termination than in the Sb termination as discussed in the previous paragraph.
In going from the Fermi level to -0.2 eV, the spectral weight at M/L point is enhanced while that around K/H point is reduced. Consequently, the difference between the Cs and Sb terminations is not so clear at -0.2 eV.

\begin{figure}[t]
\centering
\includegraphics[width=0.48\textwidth,pagebox=cropbox,clip]{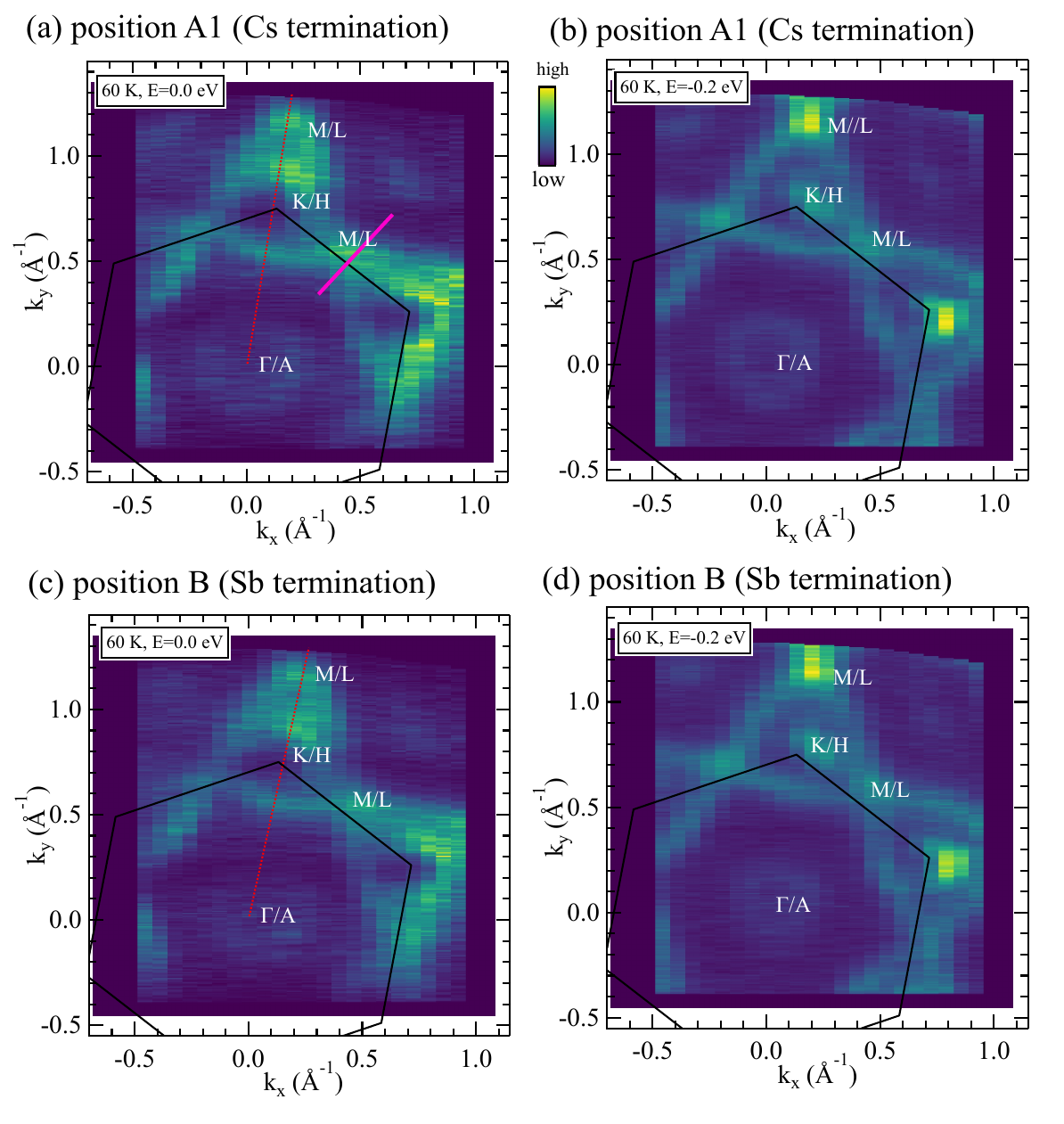}
\caption{(Color online)
(a) Fermi surface map and (b) constant energy map at -0.2 eV taken at 60 K for position A1 (Cs termination).
(c) Fermi surface map and (d) constant energy map at -0.2 eV taken at 60 K for position B (Sb termination).
The hexagonal lines indicate the Brillouin zone boundaries.
}
\end{figure}

Figure 6(a) shows a band map along the $\Gamma$-K-M line in Fig. 5(a) for position A1 (Cs termination). The observed dispersions are similar to the DFT calculations without lattice distortions as shown by the dashed and solid curves. The relatively small impact of the charge density wave on the band dispersions suggests weakness of electronic correlation. The surface sensitive ARPES results on the weak electronic correlation is consistent with the bulk-sensitive hard x-ray photoemission work by Takegami et al. \cite{Takegami2024}. 
Figure 6(b) shows a band map along the $\Gamma$-K-M line in Fig. 5(a) for position B (Sb termination). Intensity of the Sb 5$p_z$ band around $\Gamma$ is suppressed with the Sb termination. Although the spectral distribution is dramatically different between the Cs- and Sb-terminated regions as demonstrated in Figs. 2 and 3, the overall band dispersion of the Sb-terminated surface is similar to that of the Cs-terminated one. These observations are consistent with the study by Kato et al. \cite{Kato2022_2}. 

\begin{figure}[t]
\centering
\includegraphics[width=0.48\textwidth,pagebox=cropbox,clip]{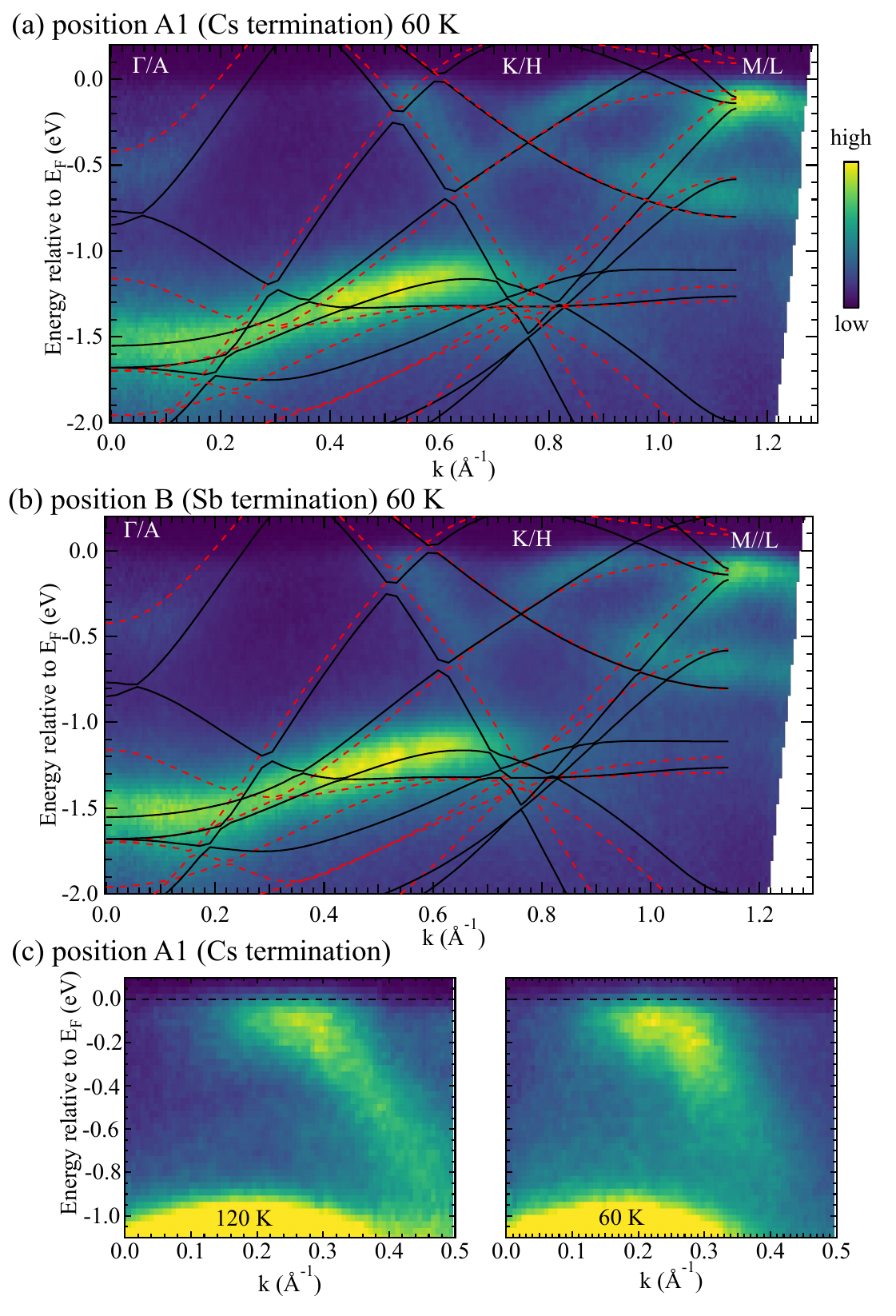}
\caption{(Color online)
Band dispersions along the $\Gamma$-K-M or A-H-L cut indicated by the dotted line in Fig. 5(a) for (a) position A1 (Cs termination) and (b) position B (Sb termination). Calculated band dispersions along $\Gamma$-K-M and A-H-L are shown by dashed and solid curves, respectively. (c) Band dispersions taken at 120 K and 60 K around M/L as indicated by the thick solid line in Fig. 5(a).}
\end{figure}

Figure 6(c) shows ARPES spectral distribution around M/L along the cut indicated by the thick solid lines in Fig. 5(a). Although the energy resolution is limited to 50 meV in the present measurement with the submicron beam spot, the intensity at M/L point increases around -0.1 eV and around -0.3 eV with cooling from 120 K to 60 K. The spectral change across the charge density wave transition is consistent with the band folding between M and L points due to the doubling of lattice constant $c$ and the gap formation at the van Hove singularity at M/L of the Cs-terminated surface as already pointed out in the previous studies \cite{Nakayama2021, Kato2022_2}. Here, it would be interesting to compare CsV$_3$Sb$_5$ with IrTe$_2$ which shows charge density wave under weak electronic correlation \cite{Pyon2012,Yang2012}. The spectral change across the transition of CsV$_3$Sb$_5$ is much smaller than the drastic change observed in IrTe$_2$ \cite{Ootsuki2012,Ootsuki2019,Nicholson2024}. The contrast between CsV$_3$Sb$_5$ and IrTe$_2$ can be associated with the effect of orbital degrees of freedom. In IrTe$_2$, Ir-Ir and Te-Te dimers are formed with the symmetry breaking of Ir 5$d$ $t_{2g}$ and Te 5$p$ orbitals. Such orbital symmetry breaking is absent in the charge density wave of CsV$_3$Sb$_5$. As for the role of electronic correlation, it is instructive to compare the charge density wave of CsV$_3$Sb$_5$ with those of $\sqrt{13}\times\sqrt{13}$ modulation in 
1T-TaS$_2$, TaSe$_2$, NbS$_2$, and NbSe$_2$ \cite{Dardel1992,Zwick1998,Pillo1999,Horiba2002,Clerc2006,Hellmann2010,Ritschel2013,Wang2017,Wang2020,Nakata2021}, in which splitting of the core level peak indicates large potential difference between Ta (or Nb) sites due to the charge density wave.
In CsV$_3$Sb$_5$, the core level peak does not change appreciably across the charge density wave transition \cite{Takegami2024}. The absence of the core level splitting under the charge density wave is similar to that observed in 1T-VSe$_2$ \cite{Terashima2003,Sereika2020,Sahoo2020}.

Figure 7(a) shows a SPEM image and extracted photoemission spectra for $\Gamma$/A point taken at 60 K. The photoemission spectra of positions A1 and B in the SPEM image are shown in the lower panel where the energy window for SPEM is shaded. In order to see the effect of charge density wave gap, the energy window is placed at the Fermi level.
The difference between the Cs- and Sb-terminated surfaces is not so clear in Fig. 7(a) compared to Fig. 2(a). However, the photoemission intensity is highly inhomogeneous in the region of Sb-terminated surface (around position B). This observation suggests that the disorder/defect of the surface Sb2 site is related to the electronic inhomogeneity. It is speculated that the Sb 5$p_z$ band is strongly affected by the disorder/defect of the Sb2 site in the Sb-terminated surface and would be relevant for the partial suppression of the 2$\times$2 and 1$\times$4 charge density waves at the Sb-terminated surface as reported by Kurtz et al. \cite{Kurtz2024} For the image shown in Fig. 7(b), a narrower region in the Cs-terminated surface is selected as indicated by the box in Fig. 7(a). Interestingly, modulation of the photoemission intensity is clearly seen although it is weaker than that for the Sb termination. The photoemission spectra for the bright and dark regions (positions A1 and A2, respectively) are extracted and shown in the lower panel. In the dark region, while the intensity near the Fermi level is reduced, the -1.5 eV peak derived from the out-of-plane Sb is enhanced. The intensity modulation suggests that electronic inhomogeneity of the Sb 5$p_z$ band in the Cs-terminated surface is likely to be associated with the atomic disorder of the out-of-plane Sb2 site which governs the hybridization between Sb 5$p_z$ and V 3$d$ at $\Gamma$/A. This observation suggests that the less inhomogeneous Cs-terminated surface is more suitable for superconducting interface. However, the noticeable inhomogeneity of the Cs-terminated surface is expected to modulate the Sb 5$p$-V 3$d$ hybridization and affect the interface.

Figure 7(c) shows a SPEM image and extracted photoemission spectra for K/H point taken at 60 K. The photoemission spectra of positions A1 and B in the SPEM image are shown in the lower panel where the energy window for SPEM is indicated. The difference between the Cs- and Sb-terminated surfaces is clearly observed at K/H point since the hybridization between V 3$d$ $xy$/$x^2-y^2$ and Sb 5$p_x$/5$p_y$ orbitals at K/H is strongly affected by the Sb termination. In the narrower image in the Cs-terminated surface displayed in Fig. 7(d), the photoemission intensity map near the Fermi level is more homogeneous than that at $\Gamma$/A point. Figure 7(e) shows a SPEM image and extracted photoemission spectra for M/L point taken at 60 K. The photoemission spectra of positions A1 and B in the SPEM image are shown in the left panel. In the narrower image of the Cs-terminated surface displayed in Fig. 7(f), the photoemission intensity near the Fermi level is also very homogeneous. As shown in the right panel, the spectral difference between positions A1 and A2 is not appreciable. This observation is consistent with the DFT prediction that the hybridization between V 3$d$ and Sb 5$p$ orbitals at M/L point is very small and, consequently, the effect of the surface disorder is limited \cite{Hu2022}.
The SPEM image at K/H in Fig. 7(d) is more inhomogeneous that that at M/L point in Fig. 7(f). This observation suggests that, even in the Cs-terminated surface, the hybridization between V 3$d$ $xy$/$x^2-y^2$ and Sb 5$p_x$/5$p_y$ orbitals is affected by the surface disorder.

\clearpage
\onecolumngrid

\begin{figure}
\centering
\includegraphics[width=0.96\textwidth,pagebox=cropbox,clip]{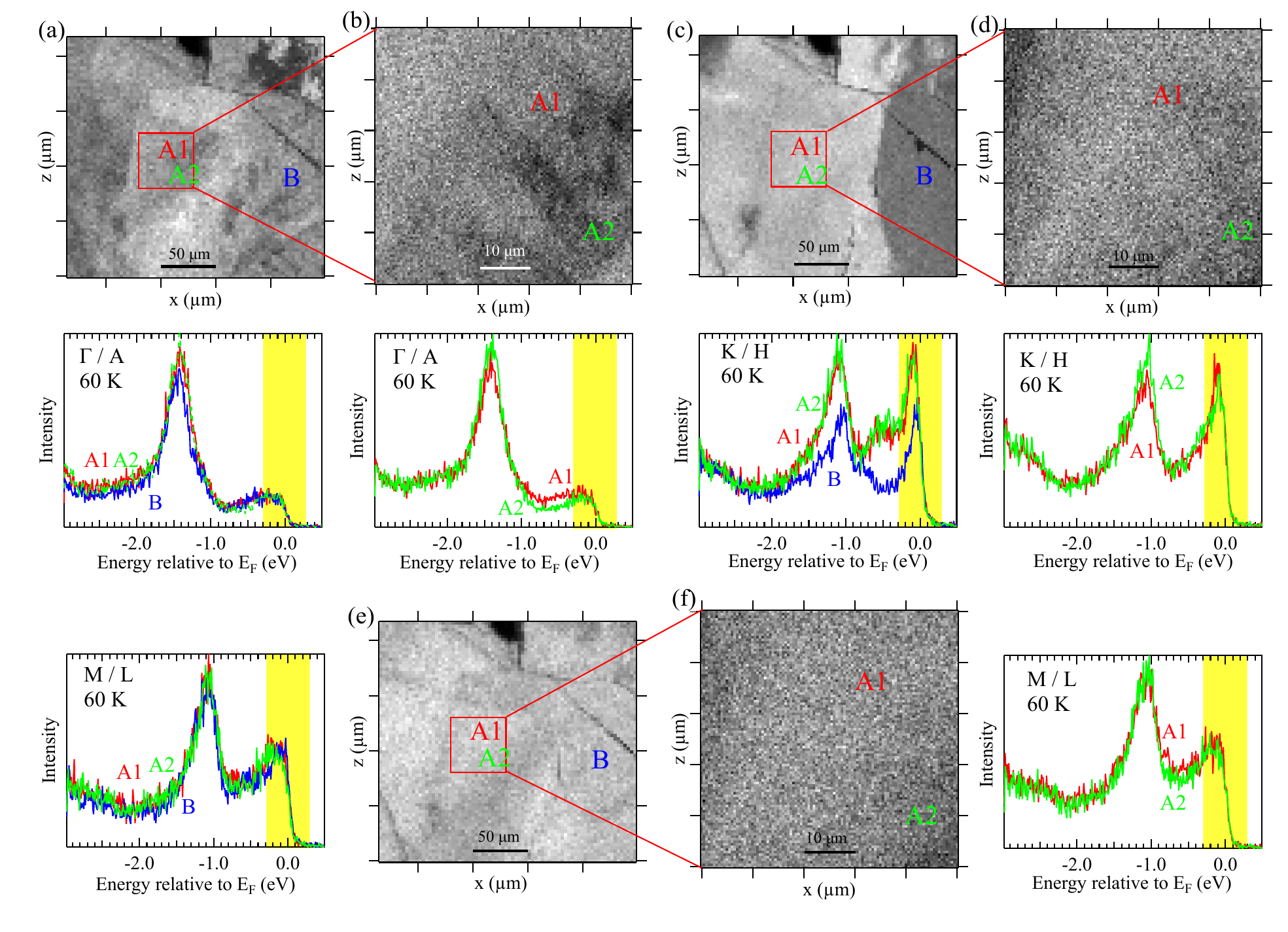}
\caption{(Color online)
(a) SPEM image and photoemission spectra for $\Gamma$/A point taken at 60 K and (b) those for a narrower region.
(c) SPEM image and photoemission spectra for K/H point taken at 60 K and (d) those for a narrower region.
(e) SPEM image and photoemission spectra for M/L point taken at 60 K and (f) those for a narrower region.
}
\end{figure}

\twocolumngrid

\section{Conclusion}
Cs- and Sb-terminated surfaces of a kagome superconductor CsV$_3$Sb$_5$ have been studied by means of SPEM. The observed band structure of the Cs-terminated surface is rather close to that of the bulk while that of the Sb-terminated one is substantially modified. The effect of the Sb-termination is substantial around K/H point where the hybridization between V 3$d$ $xy$/$x^2-y^2$ and Sb 5$p_x$/5$p_y$ orbitals is important. The SPEM results suggest that the less inhomogeneous Cs-terminated surface is more suitable for interface of kagome superconductors. However, even within the Cs-terminated region, the SPEM image at $\Gamma$/A indicates electronic inhomogeneity related to Sb 5$p$ orbitals. The electronic inhomogeneity of the Sb 5$p$ band would be driven by the atomic disorder of the out-of-plane Sb and slightly develops below the charge density wave transition temperature. Since the disorder of the out-of-plane Sb is associated with the band folding along $\Gamma$-A with the exotic charge density wave, the electronic inhomogeneity of the Cs-terminated region is affected by the transition. The electronic inhomogeneity of the Cs-terminated surface is expected to affect the Sb 5$p$-V 3$d$ hybridization at interface of the kagome superconductors.

\section*{Acknowledgements}
We thank Elettra staff for the assistance during the experimental run. This work was partially supported by Grants-in-Aid from JSPS (JP22H01172).

\end{document}